\documentclass[final,3p,times,twocolumn]{elsarticle}

\usepackage{amssymb}

\usepackage{natbib}   
\usepackage{geometry} 
\usepackage{graphicx}  
\usepackage{txfonts}   
\usepackage[colorlinks,linkcolor=blue,anchorcolor=blue,citecolor=blue,urlcolor=blue]{hyperref}
\usepackage{bm}
\usepackage[mathlines]{lineno}
\usepackage{color,soul}
\newcommand{\modR}[1] {{#1}}

\journal{Carbon}
\begin{document}


\begin{frontmatter}



\title{\hfill {\normalsize\rm Accepted in Carbon (2019)}\\
In-Plane Breathing and Shear Modes in Low-Dimensional Nanostructures}



\author[1]{Dan~Liu} %

\author[2]{Colin~Daniels}

\author[2]{Vincent~Meunier}

\author[3]{Arthur~G.~Every}

\author[1]{David Tom\'{a}nek%
\corref{cor}} %
\ead{tomanek@pa.msu.edu}

\address[1]{Physics and Astronomy Department,
            Michigan State University,
            East Lansing, Michigan 48824, USA}
\address[2]{Department of Physics,
            Applied Physics, and Astronomy,
            Rensselaer Polytechnic Institute,
            Troy, NY 12180, USA}
\address[3]{School of Physics,
            University of the Witwatersrand,
            Private Bag 3,
            2050 Johannesburg,
            South Africa}
\cortext[cor]{Corresponding author} %


\begin{abstract}
We use continuum elasticity theory to revise scaling laws for
radial breathing-like and shear-like vibration modes in quasi-2D
nanostructures including finite-width nanoribbons and finite-size
thin circular discs. Such modes can be observed spectroscopically
in corresponding nanostructures of graphene and phosphorene and
can be determined numerically by atomistic {\em ab initio} density
functional theory and classical force-field calculation. The
revised scaling laws differ from previously used expressions, some
of which display an unphysical asymptotic behavior. Apart from
model assumptions describing the effect of edge termination, the
continuum scaling laws have no adjustable parameters and display
correct asymptotic behavior. These scaling laws yield excellent
agreement with experimental and numerical results for vibration
frequencies in both isotropic and anisotropic structures as well
as useful expressions for the frequency dependence on structure
size and edge termination.
\end{abstract}




\end{frontmatter}


\section{Introduction}

Well-defined nanostructures of carbon and other elements,
including the C$_{60}$ buckyball~\cite{SmalleyC60} and other
fullerenes, carbon nanotubes~\cite{Iijima1991}, and graphene
nanoribbons~\cite{DT227} have been produced with atomic-scale
precision~\cite{Fasel2010}. Resonant Raman spectroscopy has
emerged as the most powerful method to identify and characterize
each of these nanostructures within a sample. There are extensive
theoretical and experimental
studies~\cite{{Rao1997},{Ando02},{Ferrari2004},{Maultzsch2005}}
that relate the frequency $\omega_{RBM}$ of the radial breathing
mode (RBM) of carbon fullerenes and nanotubes to their diameter.
In turn, the diameter of these nanostructures can be inferred from
Raman spectra. %

\modR{%
For finite-width 2D nanoribbons, several theoretical studies
utilized time-consuming atomistic total energy calculations and
translated frequency results to frequency-width scaling
laws~\cite{{Zhou2007},{Meunier2008},{Yamada2008},{Dong2008}} that
displayed an unphysical asymptotic behavior for wide nanoribbons.
Other studies~\cite{{Gillen09},{Gillen10},{Gillen10PRB}} obtained
the correct asymptotic behavior for wide nanoribbons using the
Brillouin zone folding approach, but could not find agreement with
experimental data in narrow nanoribbons.
}%

\modR{%
With precision rivaling that of {\em ab initio} calculations at
much lower computational effort, %
}%
RBM frequencies of near-spherical fullerenes and cylindrical
nanotubes have been obtained by representing these nanostructures
by elastic membranes characterized by 2D elastic
constants~\cite{DT255}. In the standard Voigt notation, adapted to
a 2D solid~\cite{DT255}, these yield
\begin{equation}
\omega_{C_n} = \frac{2}{d} \sqrt{\frac{2c_{11}}{\rho_{2D}}} \,,%
\label{Eq1}
\end{equation}
for the RBM frequency of a spherical C$_n$ fullerene with diameter
$d$. Similarly, the RBM frequency of a carbon nanotube (CNT) of
radius $d$ is given by
\begin{equation}
\omega_{CNT} = \frac{2}{d} \sqrt{\frac{c_{11}}{\rho_{2D}}} \,.%
\label{Eq2}
\end{equation}
The planar counterpart of a cylindrical CNT is an infinite, planar
nanoribbon of width $W$. In analogy, planar circular nanodiscs of
radius $R$ are planar counterparts of spherical fullerenes.


\modR{%
Here we present continuum elasticity theory results that translate
into revised scaling laws for radial breathing-like and shear-like
vibration modes in quasi-2D nanostructures including finite-width
nanoribbons and finite-size thin circular discs. Such modes have
been observed spectroscopically in corresponding nanostructures of
graphene and phosphorene and can be determined numerically by
atomistic {\em ab initio} density functional theory and classical
force-field calculation. The revised scaling laws differ from
previously used expressions, some of which display an unphysical
asymptotic behavior in wide nanostructures. Apart from model
assumptions describing the effect of edge termination, the
continuum scaling laws have no adjustable parameters and display
correct asymptotic behavior. These scaling laws yield excellent
agreement with experimental and numerical results for vibration
frequencies in both isotropic and anisotropic structures as well
as useful expressions for the frequency dependence on structure
size and edge termination. %
}%


\modR{%
\section{%
Analytical Scaling Laws%
}%
}

\subsection{%
\modR{%
Expressions for %
}%
Breathing-Like and Shear-Like Modes of Nanoribbons}

\begin{figure*}[t]
\includegraphics[width=2.0\columnwidth]{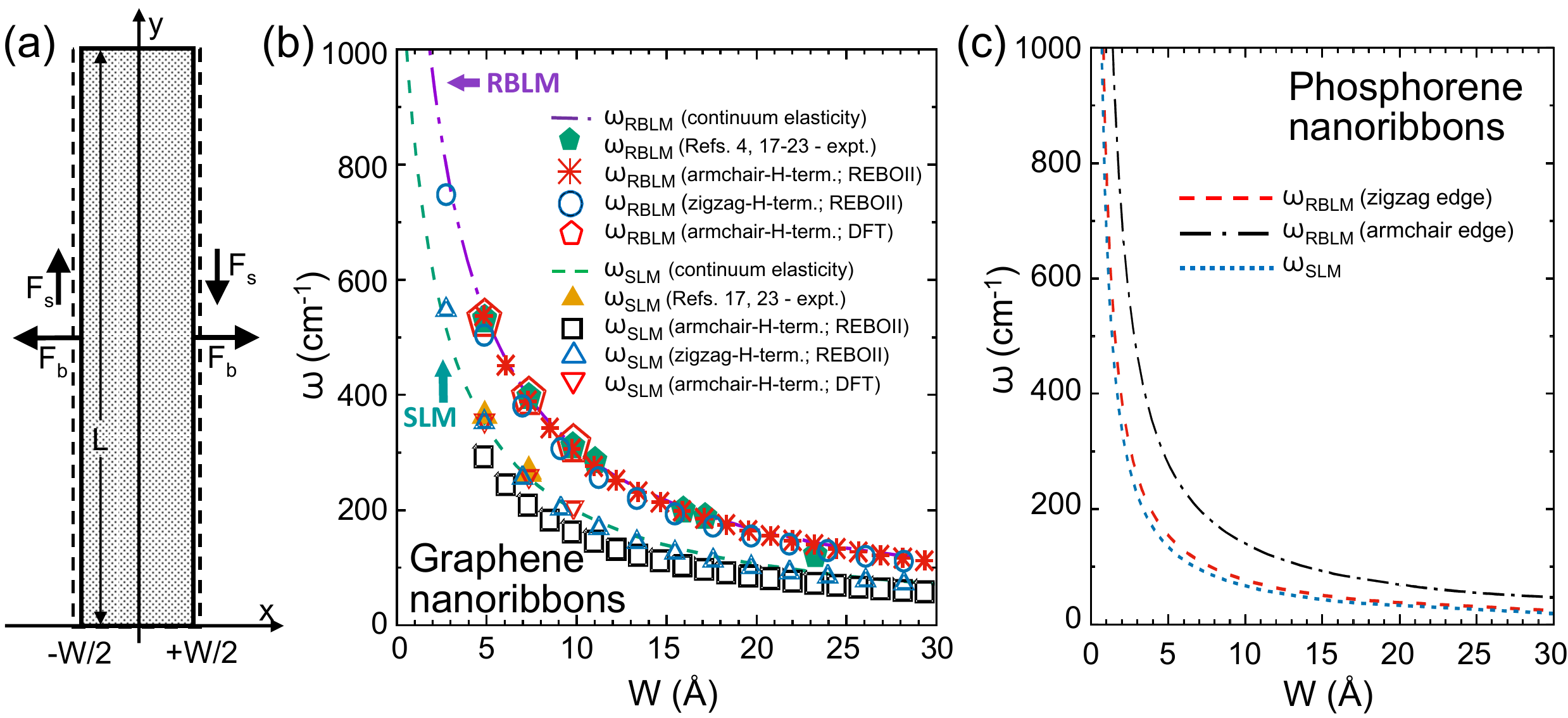}
\caption{In-plane breathing-like and shear-like modes of an
infinite planar nanoribbon of bare width $W$. %
(a) Schematic of the deformations considered here. %
(b) Numerical results for the $W$-dependence of the radial
breathing-like mode frequency $\omega_{RBLM}$ and the shear-like
mode (SLM) frequency $\omega_{SLM}$ in armchair graphene nanoribbons
(aGNRs) using the continuum elasticity and atomistic approaches. %
(c) Continuum elasticity results for the $W$-dependence of
$\omega_{RBLM}$ and $\omega_{SLM}$ in phosphorene nanoribbons cut
along the softer direction 1 or the harder direction 2. %
Experimental data are reproduced from
Refs.~\cite{%
{Fasel2010},{Ma2017},{Borin2019},{Zhao2017},{Talirz2017},%
{Llinas2017},{Chen2017},{Overbeck2019}}. %
%
} %
\label{fig1}
\end{figure*}

\modR{%
Let us first consider an infinitely long nanoribbon of width $W$
and 2D mass density ${\rho_{2D}}$, represented in
Fig.~\ref{fig1}(a), which is free of tensile strain energy. %
}%
To obtain an expression for the in-plane radial breathing-like
mode (RBLM), we first consider two longitudinal acoustic (LA)
waves propagating in opposite direction within an infinite layer
with velocity $v={\omega}/k$, producing a standing wave with
${\lambda}=2{\pi}/k$ as interference pattern. This pattern is an
array of infinitely long strips, separated by stress nodal lines
${\lambda}/2$ apart. Cutting the plane along adjacent stress nodal
lines of the standing wave, which are immobile in the plane, will
produce a ribbon of finite width $W={\lambda}/2$ that will vibrate
with the same frequency $\omega$ if edge effects can be ignored.
2D continuum elasticity theory calculations provide the
expression~\cite{DT255}
\begin{equation}
\omega_{LA}(k) = \sqrt{\frac{c_{11}}{\rho_{2D}} } ~ k%
\label{Eq3}
\end{equation}
for the LA mode of the infinite layer corresponding to the
in-plane breathing-like mode of a nanoribbon. Substituting
$k=2{\pi}/{\lambda}=2{\pi}/(2W)$ from above in Eq.~(\ref{Eq3}),
we obtain
\begin{equation}
\omega_{RBLM,0} = \frac{\pi}{W} \sqrt{ \frac{c_{11}}{\rho_{2D}} } %
\label{Eq4}
\end{equation}
for the RBLM frequency %
\modR{%
$\omega_{RBLM,0}$ that ignores any edge effects. %
A similar expression has been derived
previously~\cite{{Gillen09},{Gillen10},{Gillen10PRB}} using the
Brillouin zone folding approach based on the phonon spectrum of an
infinite monolayer. %
}%
In an anisotropic material, $c_{11}$ has to be
replaced by the elastic constant associated with the direction
normal to the nanoribbon axis.

\modR{%
In nanoribbons with bare edges, the width $W$ is defined
by the distance between atoms at opposite edges. Most experiments,
however, %
}%
are performed not on bare, but rather chemically terminated
nanoribbons. Chemical functionalization, such as H- or OH-
termination of graphene edges, changes the elastic response at the
edge. More important, it effectively increases the width $W$ and
mass of the nanoribbon by a constant amount per length of the
nanoribbon. Both latter effects lower the frequency of the RBLM
mode and can be accommodated by effectively increasing the bare
width $W$ by ${\delta}W$ in Eq.~(\ref{Eq4}). The value of
${\delta}W$ will represent changes in the edge region with respect
to the nanoribbon material. ${\delta}W$ should not change when the
nanoribbon width changes.

\modR{%
In view of these considerations, the scaling law of
Eq.~(\ref{Eq4}) should be modified to %
}%
\begin{eqnarray}
\omega_{RBLM} &=& \frac{c_{RBLM}}{W+{\delta}W}
\label{Eq5} \\%
\textrm{with}\
     c_{RBLM} &=& {\pi} \sqrt{ \frac{c_{11}}{\rho_{2D}} }
\;, \nonumber%
\end{eqnarray}
where edge effects are described by ${\delta}W$ as the only
adjustable parameter. The effect of specific edge terminations on
$\omega_{RBLM}$ will be discussed later on.

\modR{%
We note that the functional dependence of $\omega_{RBLM}$ on the
nanoribbon width $W$ in Eq.~(\ref{Eq5}) differs significantly from
the previously used
expression~\cite{{Zhou2007},{Meunier2008},{Yamada2008}}
$\omega_{RBLM}=a/\sqrt{W}+b$, where
$a{\approx}1.6{\times}10^3$~cm$^{-1}${\AA}$^{1/2}$ and
$b{\approx}{-2{\times}10^2}$~cm$^{-1}$ have been obtained by
numerical fits to observed and calculated frequencies for a finite
range of widths, with the asymptotic behavior
$\omega_{RBLM}{\approx}{-2{\times}10^2}$~cm$^{-1}$ for
$W{\rightarrow}\infty$. %
An alternative expression~\cite{Dong2008} %
$\omega_{RBLM}=c/W+d/\sqrt{W}+e$, %
with $c{\approx}1.6{\times}10^3$~cm$^{-1}${\AA}, %
$d{\approx}4{\times}10^2$~cm$^{-1}${\AA}$^{1/2}$ and %
$e=-10$~cm$^{-1}$, has been proposed subsequently for an extended
range of nanoribbon widths, with the asymptotic behavior
$\omega_{RBLM}=-10$~cm$^{-1}$ for $W{\rightarrow}\infty$. The
asymptotic behavior is clearly incorrect in both expressions,
since it should approach zero for $W{\rightarrow}\infty$.
Comparison of the dependence of $\omega_{RBLM}$ using the
different expressions is provided in the Supplemental Material
(SM)~\cite{2DRBLM19-SM}. %
}%

\modR{%
Similar to Eqs.~(\ref{Eq4}) and (\ref{Eq5}) %
}%
for the RBLM of a nanoribbon, we can describe the shear-like mode
(SLM) frequency of the same nanoribbon by~\cite{DT255}
\begin{equation}
\omega_{SLM,0} = \frac{\pi}{W} \sqrt{ \frac{c_{66}}{\rho_{2D}} } %
\label{Eq6}
\end{equation}
in analogy to the in-plane transverse acoustic (TA) mode in the
infinite layer.

\modR{%
We describe the effect of edge termination on the SLM modes in the
same way as on the RBLM mode in the expression %
}%
\begin{eqnarray}
\omega_{SLM} &=& \frac{c_{SLM}}{W+{\delta}W} \;,
\label{Eq7} \\%
c_{SLM} &=& {\pi} \sqrt{ \frac{c_{66}}{\rho_{2D}} }\;. \nonumber%
\end{eqnarray}
The value of ${\delta}W$ is the same as in Eq.~(\ref{Eq5}). The
effect of specific edge terminations on $\omega_{SLM}$ will be
discussed later on.

\begin{figure}[t]
\includegraphics[width=1.0\columnwidth]{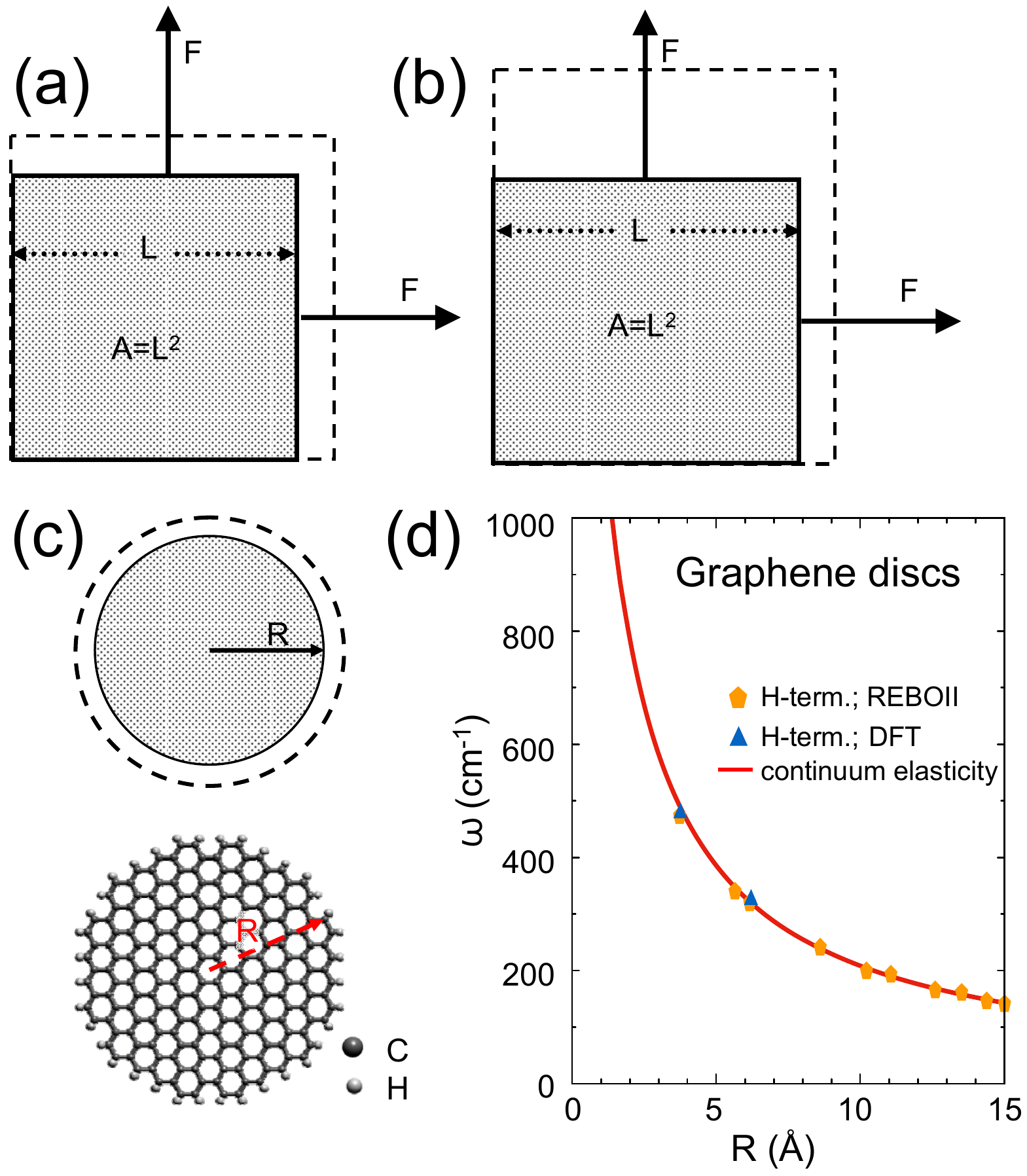}
\caption{In-plane deformation of a finite-size planar flake in the
initial shape of a square of side length $L$ for an %
(a) isotropic and %
(b) anisotropic material. %
(c) Schematic radial motion in a thin circular disc of radius $R$
    and a ball-and-stick representation of a graphene disc. %
(d) Continuum elasticity results for the $R$-dependence of
    $\omega_{RBM}$ in a graphene disc with H-terminated edges.
    The data points are results of atomistic calculations
    using the REBOII~\cite{REBOII} force-field.} %
\label{fig2}
\end{figure}


\subsection{Expression for the In-Plane Breathing Mode of
            \modR{%
            Thin Circular Discs %
            }%
            based on 2D Continuum Elasticity Theory}

Continuum elasticity theory allows to calculate the frequency of
the radial breathing mode $\omega_{RBM}$ of a solid sphere of
radius $R$, consisting of a uniform, isotropic material of mass
density $\rho$, with the help of elastic constants $C_{ij}$,
including the related bulk modulus $B$. In a similar way, the 2D
bulk modulus $\gamma$ and other 2D elastic constants $c_{ij}$
should provide a useful expression for $\omega_{RBM}$ of a massive
circular disc of radius $R$ and mass density $\rho_{2D}$, which
could be measured by Raman spectroscopy. While $\omega_{RBM}$
depends independently on both $c_{11}$ and $c_{12}$, it is
particularly sensitive to the 2D bulk modulus
$\gamma=(c_{11}+c_{12})/2$ described below.

Extensive atomistic calculations of the 2D bulk modulus $\gamma$
of graphene and a number of other isotropic quasi-2D solids have
been reported previously~\cite{Chetty12}. The 2D bulk modulus,
also known as ``membrane stretching modulus'' or ``area-stretching
elastic constant'', is a measure of the elastic resistance of a 2D
solid to change in the area under isotropic line force $F$ applied
in the two in-plane directions. We found one available expression
for $\gamma$, published in Ref.~\cite{Chetty12}, to be incorrect,
apparently due to the faulty assumption that an isotropic stress
results in an isotropic strain.
In the following we derive the correct expression for the 2D bulk
modulus $\gamma$ of an anisotropic solid.


\subsubsection{Continuum Elasticity Expression for the
               2D Bulk Modulus of an Anisotropic Material}

Consider a finite-size square object that is cut out of a
generally anisotropic 2D solid and subject to uniform tensile
stress $P={\sigma}_{11}$=${\sigma}_{22}$ in the in-plane
directions, as shown in Fig.~\ref{fig2}. The stress $P=F/L$,
caused by an external force $F$ acting on side length $L$, is the
2D counterpart of uniform pressure, and causes a fractional change
in area, or areal strain
${\delta}A/A=\epsilon_{11}+\epsilon_{22}$. There are no shear
stresses, and any shear strain ${\epsilon}_{12}$ does not
contribute to a change in the area, and so is not shown in
Fig.~\ref{fig2}. In the standard Voigt notation, adapted to a 2D
solid~\cite{DT255}, the stress-strain relationship takes the form
\begin{eqnarray}
\left( \begin{array}{ccc}
 {\epsilon}_{11}\\
 {\epsilon}_{22}\\
 {\epsilon}_{12}
       \end{array}
\right) = %
\left( \begin{array}{ccc}
 s_{11} & s_{12} & s_{16} \\
 s_{12} & s_{22} & s_{26} \\
 s_{16} & s_{26} & s_{66}
       \end{array}
\right)
\left( \begin{array}{ccc}
 P\\
 P\\
 0
       \end{array}
\right) \,,%
\label{Eq8}
\end{eqnarray}
where $s_{ij}$ are the 2D compliance constants, which form a
symmetric matrix. It follows that the areal extensibility $k$, the
2D counterpart of 3D compressibility, is given by
\begin{equation}
k = \frac{{\delta}A}{A}{\cdot}\frac{1}{P} %
=\frac{{\epsilon}_{11}+{\epsilon}_{22}}{P} %
=s_{11}+s_{22}+2s_{12} \,.%
\label{Eq9}
\end{equation}
The 2D bulk modulus $\gamma$ is the inverse of the extensibility,
so that
\begin{equation}
\gamma
=\frac{1}{k}%
=\frac{1}{s_{11}+s_{22}+2s_{12}} \,.%
\label{Eq10}
\end{equation}
For 2D structures, the elastic stiffness constants $c_{ij}$ are in
more common use than the compliance constants~\cite{DT255}, so it
is useful to be able to express the bulk modulus in terms of
stiffness constants. In principle this is straightforward since
the stiffness and compliance matrices are the inverse of each
other. For general anisotropy, numerical inversion seems the most
viable option. Of more practical importance, for square, hexagonal
and rectangular structures, $s_{16}$=$s_{26}$=0 and
$c_{16}$=$c_{26}$=0, and the required inverted compliance
constants are readily obtained as
\begin{eqnarray}
s_{11}&=&\frac{c_{22}}{c_{11}c_{22}-c_{12}^2} \,, \nonumber \\
s_{12}&=&\frac{-c_{12}}{c_{11}c_{22}-c_{12}^2} \,, \nonumber \\
s_{22}&=&\frac{c_{11}}{c_{11}c_{22}-c_{12}^2} \,.%
\label{Eq11}
\end{eqnarray}
Inserting these expressions in Eq.~(\ref{Eq10}), one arrives at
\begin{equation}
\gamma=\frac{c_{11}c_{22}-c_{12}^2}{c_{11}+c_{22}-2c_{12}} \,.
\label{Eq12}
\end{equation}
For an isotropic solid with $c_{11}=c_{22}$, Eq.~(\ref{Eq12})
reduces to
\begin{equation}
\gamma = \frac{1}{2}(c_{11}+c_{12}) \,. %
\label{Eq13}
\end{equation}
%
%


\subsubsection{Continuum Elasticity Description of In-Plane
               Vibrations in Circular Discs}


In-plane vibrations of circular discs have been extensively
discussed in the literature, including Ref.~\cite{Bashmal2010} and
references cited therein.
The fundamental frequency is readily obtained from the equation
for the radial modes of an isotropic cylinder, which can be found
e.g. in Rose's book~\cite{Rose-book}. Setting $k$ along the axis
to zero and dividing the right-hand side by 2, the cylinder
equation reduces to
\begin{equation}
J_1\left(\frac{{\omega}_{RBM}R}{v_{LA}}\right)
= \left(\frac{{\omega}_{RBM}Rv_{LA}}{2v_{TA}^2}\right) %
J_0\left(\frac{{\omega}_{RBM}R}{v_{LA}}\right) \,,%
\label{Eq14}
\end{equation}
where $J_0$ and $J_1$ are Bessel functions and $R$ is the radius
of the cylinder. This equation applies equally to the radial modes
of a thin disc, requiring only that
$v_{LA}=\sqrt{c_{11}/\rho_{2D}}$ be treated as the 2D longitudinal
acoustic velocity and $v_{TA}=\sqrt{c_{66}/\rho_{2D}}$ as the
transverse acoustic velocity. Remembering that
$c_{66}=(c_{11}-c_{12})/2$ in an isotropic 2D
material~\cite{DT255} we consider here, we can further define
$x={\omega}_{RBM}R/v_{LA}$ and rewrite Eq.~(\ref{Eq14}) as
\begin{equation}
J_1(x)=CxJ_0(x) \,. %
\label{Eq15}
\end{equation}
Here, $C=c_{11}/(2\gamma)(c_{11}+c_{12})/(c_{11}-c_{12})=
c_{11}/(c_{11}-c_{12})$ is a constant depending on the elastic
constants of the material. According to its definition, the root
of Eq.~(\ref{Eq15}) is related to the fundamental frequency of a
circular isotropic disc by $x={\omega_{RBM}}R
\sqrt{\rho_{2D}/c_{11}}$, so that
\begin{equation}
{\omega_{RBM}} = \frac{x}{R}\sqrt{\frac{c_{11}}{\rho_{2D}}}\,. %
\label{Eq16}
\end{equation}
The %
\modR{%
general %
}%
solution of Eq.~(\ref{Eq15}) has to be obtained numerically for a
given material and provides both the fundamental frequency and the
overtones.


In the following, we will apply the above expressions for the
different vibration mode frequencies to specific nanostructures,
in particular nanoribbons and
\modR{%
thin circular discs %
}%
formed of 2D graphene and phosphorene.


\modR{%
\section{%
Numerical Results for Specific Materials and Structures %
}%
}%

\subsection{%
Nanoribbons of Graphene and Phosphorene}


Our results for $\omega_{RBLM}$ and $\omega_{SLM}$ for armchair
graphene nanoribbons (aGNRs) are presented in Fig.~\ref{fig1}(b).
Graphene is isotropic and characterized by~\cite{DT255}
$c_{11}=c_{22}=352.6$~N/m, $c_{12}=59.6$~N/m, $c_{66}=146.5$~N/m,
and $\rho_{2D}=0.743{\times}10^{-6}$~kg/m$^2$. The width of bare
$N$-aGNRs is given by~\cite{DT227} $W=(N-1){\times}1.23$~{\AA} and
the width of $N$-zGNRs with a zigzag edge is given by
$W=(2.13N-1.42)$~{\AA}.

\begin{figure}[t]
\includegraphics[width=0.8\columnwidth]{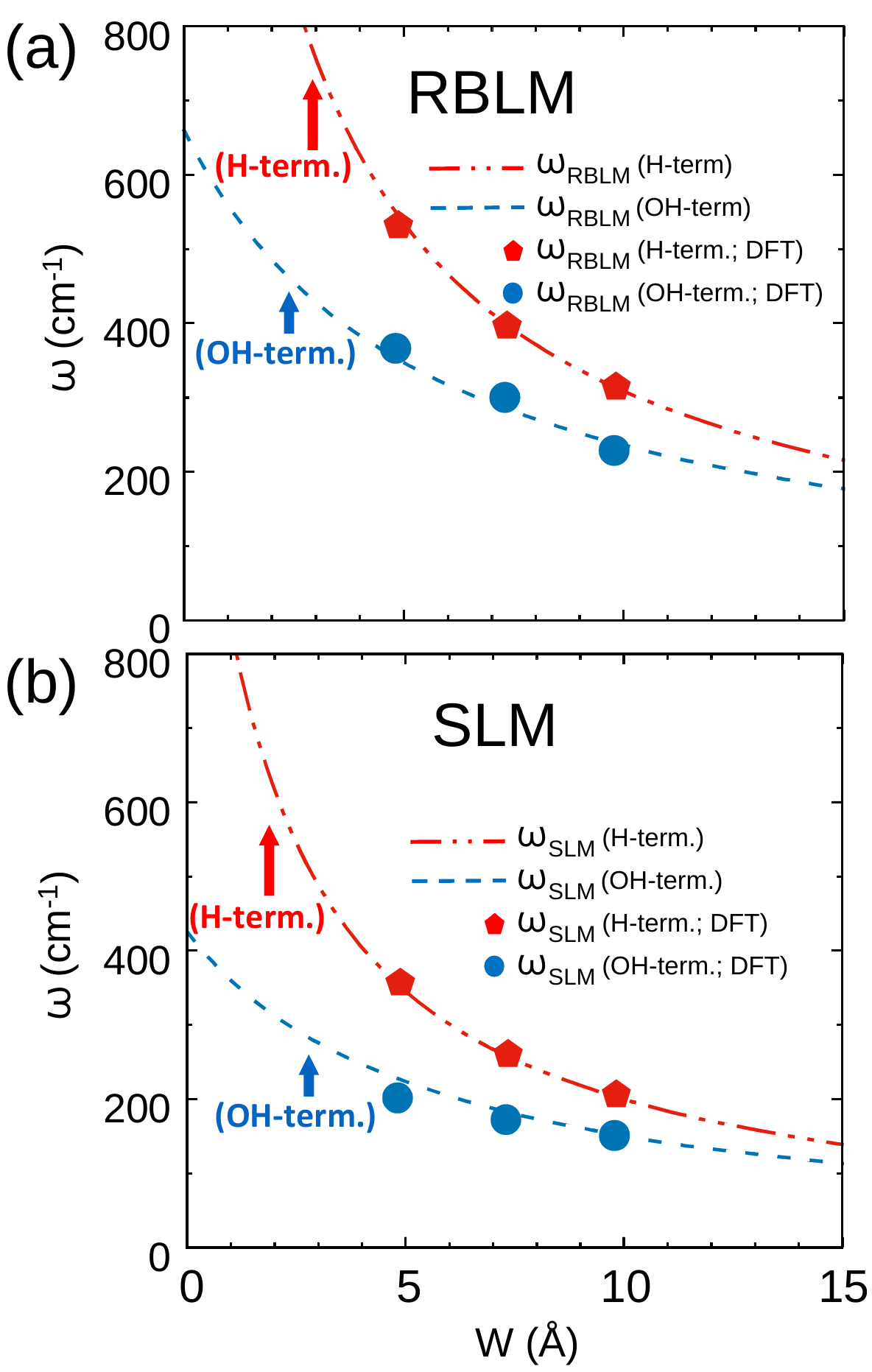}
\caption{Effect of edge termination on vibration modes in GNRs.
Results for %
(a) $\omega_{RBLM}$ and %
(b) $\omega_{SLM}$ %
obtained using Eq.~(\ref{Eq5}) are presented for edge termination
by either lighter H- or heavier -OH groups. %
} %
\label{fig3}
\end{figure}

Whereas bare edges are known to reconstruct and change their
elastic response, chemical termination adds width and mass to the
edge. Graphene edges are often terminated by H- or OH- groups in
aqueous environment~\cite{DT274}. Considering only the width
change in case of hydrogen-terminated GNRs, we may expect
${\delta}W{\approx}2.2$~{\AA} in view of the typical C-H bond
length of $1.1$~{\AA} at the edges. As seen in Fig.~\ref{fig3}, we
found that all data for $\omega_{RBLM}$ in H-terminated GNRs,
whether obtained spectroscopically or by atomistic calculations,
could be reproduced accurately by continuum elasticity theory
Eqs.~(\ref{Eq5}) and (\ref{Eq7}) using the value
${\delta}W=1.8$~{\AA} for H-termination, which agrees in the order
of magnitude with our estimate. For OH-terminated edges, continuum
elasticity calculations using ${\delta}W=5.5$~{\AA} reproduce well
numerical results for $\omega_{RBLM}$ and $\omega_{SLM}$ as seen
in Fig.~\ref{fig3}. Clearly, the higher mass of the OH-
termination is reflected in a larger value of ${\delta}W$.
Due to the isotropy of graphene, the vibration frequencies are the
same for armchair and zigzag nanoribbons of the same width, as
confirmed by results of our atomistic calculations %
\modR{ %
presented in Fig.~\ref{fig1}(b). %
}%

Unlike graphene, phosphorene is anisotropic and characterized
by~\cite{DT255} $c_{11}=24.4$~N/m, $c_{22}=94.6$~N/m,
$c_{12}=7.9$~N/m, $c_{66}=22.1$~N/m, and
$\rho_{2D}=1.34{\times}10^{-6}$~kg/m$^2$. %
\modR{%
This anisotropy is reflected in %
}%
our results for $\omega_{RBLM}$ and $\omega_{SLM}$ for armchair
and zigzag phosphorene nanoribbons, presented in
Fig.~\ref{fig1}(c).


\subsection{%
\modR{%
Thin Circular Discs %
}%
of Graphene}

Using Eq.~(\ref{Eq13}) for isotropic media and the above-mentioned
elastic constants for graphene~\cite{DT255}, we obtain the value
$\gamma=206.1$~N/m for the 2D bulk modulus of graphene, in
agreement with previously published results~\cite{Chetty12}. With
$C=c_{11}/(c_{11}-c_{12})=352.6/(352.6-59.6)=1.203$, the solution
of Eq.~(\ref{Eq15}) is $x=1.963$.
The dependence of the fundamental frequency $\omega_{RBM}$ in
\modR{%
thin circular discs %
}%
on their radius $R$ is then given by Eq.~(\ref{Eq16}) and
presented in Fig.~\ref{fig2}(d). %
\modR{%
Atomic displacements during the radial breathing modes of these
discs are displayed in the SM~\cite{2DRBLM19-SM}. %
}%

Results of atomistic calculations for hydrogen-terminated discs
based on DFT and the REBOII force field~\cite{REBOII} are compared
to continuum results in Fig.~\ref{fig2}(d). General agreement
between continuum and atomistic results provides strong support
for the universal nature of Eq.~(\ref{Eq16}) to correctly
represent $\omega_{RBM}$ of
\modR{%
thin circular discs. %
}%
Similar to bare and chemically terminated nanoribbons, the narrow
region near the edge is not represented well by continuum
elasticity theory due to changes in width, mass distribution and
local elastic behavior. Like in the case of nanoribbons, we
accommodate these effects in a single parameter ${\delta}R$ that
modifies the radius $R$ %
\modR{%
of the thin circular disc. %
}
We found good agreement between atomistic and continuum results
using ${\delta}R=0.9$~{\AA} and note that this value agrees with
${\delta}W/2$ used earlier for hydrogen-terminated nanoribbons.



\section{Discussion}

We found the accuracy of the 2D elasticity approach and its
ability to correctly represent vibration modes down to the
nanometer scale to be impressive, but not completely unexpected in
view of published results for acoustic, in particular flexural
modes of 2D materials and systems like carbon fullerenes and
nanotubes~\cite{DT255}. More important in our view is the fact
that this approach provides a physically motivated expression for
RBLM and SLM modes in nanoribbons in Eqs.~(\ref{Eq5}) and
(\ref{Eq7}), which differs from the established form and does not
suffer from an asymptotically incorrect behavior for ultra-wide
nanoribbons. In addition, by relating the frequency to intrinsic
elastic properties of the material, the provided expressions offer
a wide range of applicability for
\modR{%
2D nanostructures of any layered material. %
}%

Furthermore, even though many 2D materials such as graphene are
isotropic, some are not, with black phosphorene being a notable
example. As seen in Fig.~\ref{fig1}(c), the vibration frequencies
of phosphorene nanoribbons depend sensitively on whether their
edges are along the zigzag or the armchair direction, and this
behavior can be described by the same continuum elasticity theory.

Chemical termination of edges poses, of course, a problem for the
continuum description of finite-size objects. Depending on the
termination type, the edge region will have a different mass
distribution, different elastic behavior, and will add a nonzero
width ${\delta}W$ to finite-size objects. We believe that the
value of ${\delta}W$ in nanoribbons and the corresponding value
${\delta}R={\delta}W/2$ in
\modR{%
thin circular discs, %
}%
which
we used in our study, describes adequately the combined effect of
elastic softening near the edge and, in case of chemical
termination, an increase of the width and mass of the
nanostructure.
We should also note that the effect of chemical termination should
be different for terminating H-, O- and OH-groups.

One of the chemical edge termination effects, namely the addition
of atomic masses at the edge, can be treated analytically, as
shown in the Appendix. We find that in many cases, including
nanoribbons and
\modR{%
thin circular discs, %
}%
keeping ${\delta}W$ and ${\delta}R$ as an adjustable parameter
that is independent of $W$ or $R$ provides satisfactory results
especially for wide nanostructures. In systems, where the
termination size is similar to the size of the nano-object,
continuum elasticity treatment loses its justification.

Our calculations are all for natural normal modes of
nanostructures, whether nanoribbons or
\modR{%
thin circular discs. %
}%
We wish to note that these vibrations differ from forced modes
that may be induced by applying strain in a particular way. A
nanoribbon may be stretched uniformly normal to its axis by
applying constant 2D tensile strain along the edges. Releasing
this strain results in a soft vibration mode with a frequency
higher by the factor $\sqrt{12}/\pi$ than the eigenmode described
by Eq.~(\ref{Eq4}). This reveals that this particular forced mode
is not an eigenmode, but rather a mixed mode of the nanoribbon.


\section{Summary and Conclusions}

We have used 2D continuum elasticity theory and atomistic
calculations to determine in-plane radial breathing-like and
shear-like vibration modes of low-dimensional nanostructures
including finite-width nanoribbons and finite-size %
\modR{%
thin circular discs %
}%
of graphene and phosphorene. These vibrations can be observed by
Raman spectroscopy and used to characterize the sample.
Vibrational modes are sensitive not only to shape and mass
density, but also to anisotropy in the elastic behavior. %
\modR{%
We have derived revised scaling laws that differ from previously
used expressions, some of which display an unphysical asymptotic
behavior in wide nanostructures. Apart from model assumptions
describing the effect of edge termination, the continuum scaling
laws have no adjustable parameters and display correct asymptotic
behavior. These scaling laws yield excellent agreement with
experimental and numerical results for vibration frequencies in
both isotropic and anisotropic structures as well as useful
expressions for the frequency dependence on structure
size and edge termination. %
}%


\section{Appendix}
\setcounter{equation}{0}
\renewcommand{\theequation}{A\arabic{equation}}

\subsection{
\modR
{%
Description of %
}%
Nanoribbons with Massive Edge Termination}

An alternative way to describe the effect of edge termination is
to rigidly attach a line of constant mass density both edges. To
see its effect, let us first consider the nanoribbon of width $W$
in Fig.~\ref{fig1} that lies in the $x-y$ plane, with its axis
aligned with the $y-$direction.
Breathing modes of the ribbon, which are LA waves with ${\bf{k}}$
normal to the length of the ribbon, can be classified as symmetric
or anti-symmetric with respect to the midline of the ribbon. Both
are required to satisfy the wave equation for the medium and the
boundary conditions at the edges. In the case of symmetric modes,
for which the motions of mass elements in the nanoribbon are
related by $u(x,t)=-u(-x,t)$, solutions of the wave equation take
the form
\begin{equation}
u(x,t) = A \sin(kx) e^{i{\omega}t} %
\label{EqA1}
\end{equation}
%
with $\omega$ and $k=|{\bf{k}}|$ related by the dispersion
relation
\begin{equation}
\omega = \sqrt{\frac{c_{11}}{\rho_{2D}} } ~ k \,, %
\label{EqA2}
\end{equation}
which is identical to Eq.~(\ref{Eq3}). In the case of
antisymmetric modes for which $u(x,t)=u(-x,t)$, the form of the
wave equation changes to
\begin{equation}
u(x,t) = A \cos(kx) e^{i{\omega}t} %
\label{EqA3}
\end{equation}
with the same dispersion relation (\ref{EqA2}). We now consider a
line of constant mass density $\mu$ representing mass per length
to be rigidly connected to the edges at $x=-W/2$ and $x=+W/2$.

To determine $k$ and $\omega$, we have to impose the boundary
condition
\begin{equation}
\sigma_x({\pm}\frac{W}{2}) = \mu a_x \,, %
\label{EqA4}
\end{equation}
where $\sigma_x({\pm}{W}/{2})$ is the tensile stress at
${\pm}{W}/{2}$ and $a_x$ is the acceleration. This leads to
\begin{equation}
c_{11}\frac{\partial u}{\partial x} %
= - \mu \frac{\partial^2u}{\partial t^2} %
\label{EqA5}
\end{equation}
at the edges $x={\pm}W/2$, where the negative sign indicates that
the stretch pulls the mass at the edge inward. Inserting
expression (\ref{EqA1}) into Eq.~(\ref{EqA5}) for symmetric modes
leads to
\begin{equation}
k c_{11}\cos\left(k\frac{W}{2}\right) %
= \mu \omega^2 \sin \left(k\frac{W}{2}\right)\,. %
\label{EqA6}
\end{equation}
To eliminate the quantity $k$, we insert expression (\ref{EqA2})
into Eq.~(\ref{EqA6}) and obtain
\begin{equation}
{\omega} c_{11} \sqrt{\frac{\rho_{2D}}{c_{11}}}  %
\cos \left(\omega \frac{W}{2} \sqrt{\frac{\rho_{2D}}{c_{11}}} \right) %
= \mu \omega^2 %
\sin \left(\omega \frac{W}{2} \sqrt{\frac{\rho_{2D}}{c_{11}}} \right) \,. %
\label{EqA7}
\end{equation}
This can be simplified to the transcendental equation
\begin{equation}
\tan \left(\omega \frac{W}{2} \sqrt{\frac{\rho_{2D}}{c_{11}}} \right) %
= \frac{c_{11}}{\mu\omega} \sqrt{\frac{\rho_{2D}}{c_{11}}} \,. %
\label{EqA8}
\end{equation}
In the limiting case of no additional mass at the edge or $\mu=0$,
the right-hand side of Eq.~(\ref{EqA8}) diverges and the argument
of the tangent-function becomes $\pi/2$. The fundamental frequency
is given by
\begin{equation}
\omega \frac{W}{2} \sqrt{\frac{\rho_{2D}}{c_{11}}} = \frac{\pi}{2} \,, %
\label{EqA9}
\end{equation}
which translates to Eq.~(\ref{Eq4}) for $\omega_{RBLM}$. In the
limiting case of an infinitely heavy edge with
$\mu{\rightarrow}\infty$, the right-hand side of Eq.~(\ref{EqA8})
vanishes. For the fundamental frequency we then obtain
\begin{equation}
\omega \frac{W}{2} \sqrt{\frac{\rho_{2D}}{c_{11}}} = \pi \,. %
\label{EqA10}
\end{equation}
The calculation for antisymmetric modes follows along similar
lines and yields
\begin{equation}
\cot \left(\omega \frac{W}{2} \sqrt{\frac{\rho_{2D}}{c_{11}}} \right) %
= - \frac{c_{11}}{\mu\omega} \sqrt{\frac{\rho_{2D}}{c_{11}}} \,. %
\label{EqA11}
\end{equation}
The limiting cases are $\mu=0$, described by Eq.~(\ref{EqA10}),
and $\mu\rightarrow\infty$, described by Eq.~(\ref{EqA9}).

\section*{Acknowledgments}

DL and DT acknowledge partial support by the NSF/AFOSR EFRI 2-DARE
grant number EFMA-1433459. DT also acknowledges support by the
Mandelstam Institute for Theoretical Physics (MITP) and the Simons
Foundation, award number 509116.


\end{document}


\title{{\normalsize Supporting Information for:}\\
       In-Plane Breathing and Shear Modes in Low-Dimensional Nanostructures}

\author{Dan~Liu}
\affiliation{Physics and Astronomy Department,
             Michigan State University,
             East Lansing, Michigan 48824, USA}

\author{Colin~Daniels}
\affiliation{Department of Physics,
             Applied Physics, and Astronomy,
             Rensselaer Polytechnic Institute,
             Troy, NY 12180, USA}

\author{Vincent~Meunier}
\affiliation{Department of Physics,
             Applied Physics, and Astronomy,
             Rensselaer Polytechnic Institute,
             Troy, NY 12180, USA}

\author{Arthur~G.~Every}
\affiliation{School of Physics,
             University of the Witwatersrand,
             Private Bag 3,
             2050 Johannesburg,
             South Africa}

\author{David~Tom\'anek}
\email[E-mail: ]{tomanek@pa.msu.edu}
\affiliation{Physics and Astronomy Department,
             Michigan State University,
             East Lansing, Michigan 48824, USA}

\date{\today}


\maketitle


\renewcommand\thesubsection{\Alph{subsection}}
\setcounter{subsection}{0} %
\renewcommand{\theequation}{S\arabic{equation}}
\setcounter{equation}{0} %
\renewcommand{\thefigure}{S\arabic{figure}}
\setcounter{figure}{0} %
\renewcommand{\thevideo}{S\arabic{video}}
\setcounter{video}{0} %

\section{%
Comparison between scaling laws for the radial-like breathing mode
in graphene nanoribbons}

In the main manuscript,  we have derived a
scaling law for $\omega_{RBLM}$ as a function of nanoribbon width
$W$ based on continuum elasticity theory for 2D systems
\begin{eqnarray}
\omega_{RBLM} &=& \frac{c_{RBLM}}{W+{\delta}W}
\label{EqS1} \\%
\textrm{with}\ \ \
     c_{RBLM} &=& {\pi} \sqrt{ \frac{c_{11}}{\rho_{2D}} }
\;. \nonumber%
\end{eqnarray}
For graphene nanoribbons (GNRs), we use~\cite{DT255}
$c_{11}=352.6$~N/m, and $\rho_{2D}=0.743{\times}10^{-6}$~kg/m$^2$.
The effect of edge termination, which may vary from sample to sample, is
represented by the parameter ${\delta}W$. We found that
${\delta}W=1.8$~{\AA} describes H-termination properly.

We note that the functional dependence of $\omega_{RBLM}$ on the
nanoribbon width $W$ in Eq.~(\ref{EqS1}) differs significantly
from previously used scaling laws covering the entire range from
narrow to wide GNRs. Several authors have used the
expression~\cite{{Zhou2007},{Meunier2008},{Yamada2008}}
\begin{equation}
\omega_{RBLM} = \frac{a}{\sqrt{W}} + b \;. %
\label{EqS2}
\end{equation}
The values of the parameters $a$ and $b$ for GNRs are very similar
in References~\onlinecite{{Zhou2007},{Meunier2008},{Yamada2008}}.
Parameters obtained from fits to density functional theory (DFT)
calculations of Ref.~\onlinecite{Meunier2008} are %
$a=1667.9$~cm$^{-1}${\AA}$^{1/2}$ and %
$b=-210.2$~cm$^{-1}$. %
With these parameters, $\omega_{RBLM}$ becomes negative and
unphysical in wide GRNs.

\begin{figure}[b]
\includegraphics[width=0.8\columnwidth]{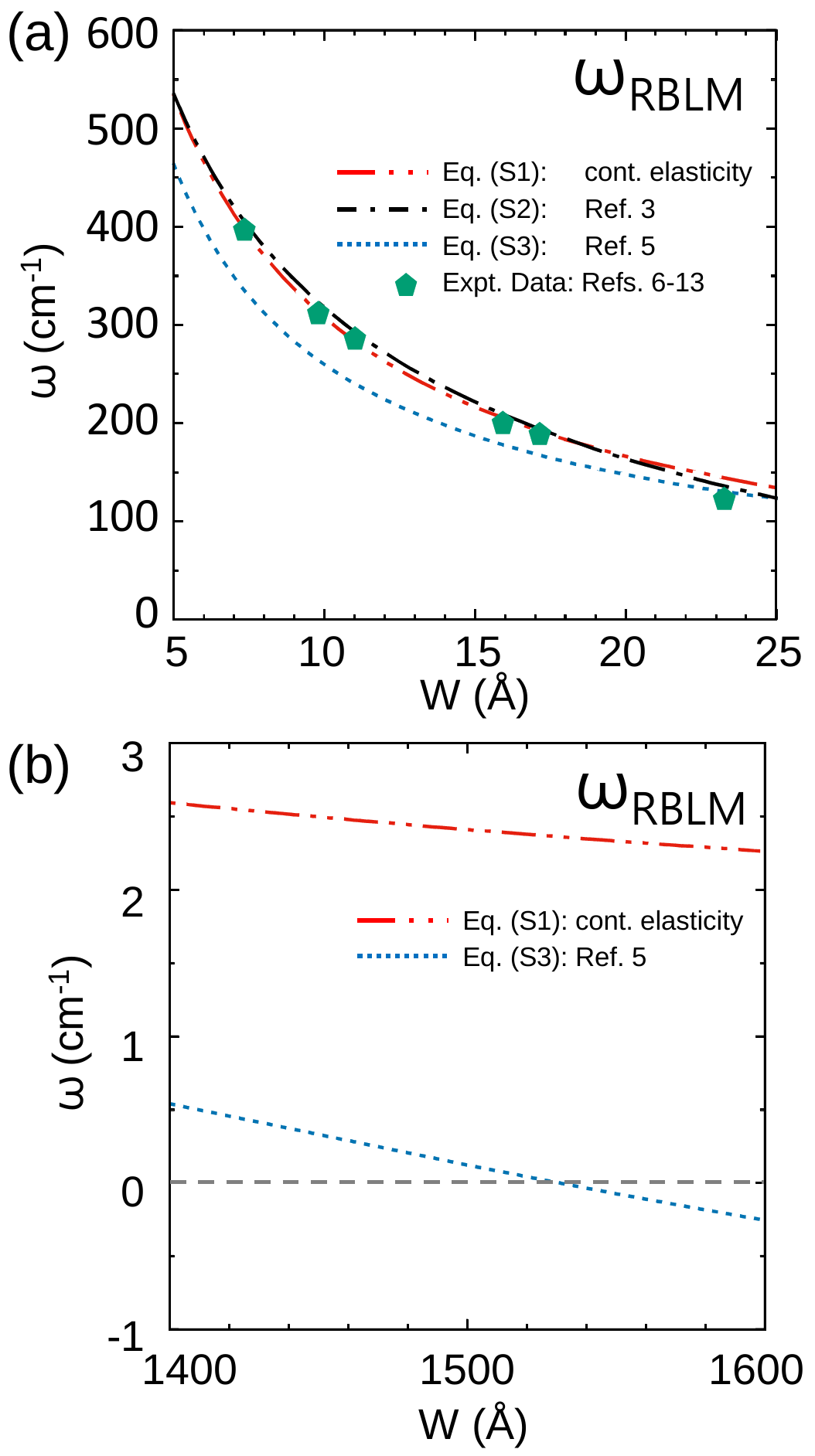}
\caption{Dependence of $\omega_{RBLM}$ on the GNR width $W$
according to expressions in %
Eqs.~(\ref{EqS1}), %
(\ref{EqS2}) and %
(\ref{EqS3}). %
Numerical results are presented for $W$ in the range of %
(a) $5-25$~{\AA} and %
(b) $1400$~{\AA}$-1600$~{\AA}.
} %
\label{figS1}
\end{figure}

An alternative expression~\cite{Dong2008} %
\begin{equation}
\omega_{RBLM} = \frac{c}{W} + \frac{d}{\sqrt{W}} + e %
\label{EqS3}
\end{equation}
has been proposed to alleviate the incorrect behavior of
$\omega_{RBLM}$ in Eq.~(\ref{EqS2}) for wide GNRs. The parameters
used in Ref.~\onlinecite{Dong2008} for GNRs are
$c=1584.24$~cm$^{-1}${\AA}, %
$d=351.98$~cm$^{-1}${\AA}$^{1/2}$ and %
$e=-10.00$~cm$^{-1}$. Yet also with this scaling law,
$\omega_{RBLM}$ eventually becomes negative in ultra-wide GNRs.

The dependence of $\omega_{RBLM}$ in hydrogen-terminated GNRs of
width $W$, described by Eqs.~(\ref{EqS1}), (\ref{EqS2}) and
(\ref{EqS3}), is compared for different width ranges in
Fig.~\ref{figS1} and experimental data.

As seen in Fig.~\ref{figS1}(a), all expressions correctly predict
a monotonic decrease of $\omega_{RBLM}$ with increasing GNR width
$W$ in narrow GNRs. There is, however, a quantitative difference
of up to ${\lesssim}20\%$ between the results for narrow GNRs with
$5{\lesssim}W{\lesssim}25$~{\AA}. Experimental
data~\cite{{Fasel2010},{Ma2017},{Overbeck2019},%
{Borin2019},{Zhao2017},{Talirz2017},%
{Llinas2017},{Chen2017}} agree best with the scaling laws in
Eqs.~(\ref{EqS1}) and (\ref{EqS2}).

Qualitative differences between the different expressions become
more obvious in wide GNRs with
$1400$~{\AA}${\lesssim}W{\lesssim}1600$~{\AA} according to
Fig.~\ref{figS1}(b). Whereas $\omega_{RBLM}$ remains positive and
correctly approaches zero for $W{\rightarrow}\infty$ according to
Eq.~(\ref{EqS1}), the frequency becomes negative for $W>63$~{\AA}
according to Eq.~(\ref{EqS2}) and for $W>1530$~{\AA} according to
Eq.~(\ref{EqS3}). Asymptotically, for $W{\rightarrow}\infty$,
$\omega_{RBLM}=-210.2$~cm$^{-1}$ according to Eq.~(\ref{EqS2}) and
$\omega_{RBLM}=-10$~cm$^{-1}$ according to Eq.~(\ref{EqS3}).

A $1/W$ functional dependence of $\omega_{RBLM}$ with the correct
asymptotic behavior $W{\rightarrow}\infty$ has been discussed
previously~\cite{{Dong2008},{Gillen09},{Gillen10},{Gillen10PRB}}
with the conclusion that it only may reproduce observed data in
"not too narrow nanoribbons"~\cite{Gillen10PRB}. We also note that
the zone-folding approach used in
Refs.~\onlinecite{{Gillen09},{Gillen10},{Gillen10PRB}} can not
easily accommodate edge effects in narrow GNRs. Among the proposed
functional dependencies of $\omega_{RBLM}$ on $W$, only
Eq.~(\ref{EqS1}) covers the entire range from narrow to wide
nanoribbons.

\section{Atomic displacements in radial breathing modes of
circular graphene discs}

\begin{figure}[h]
\includegraphics[width=1.0\columnwidth]{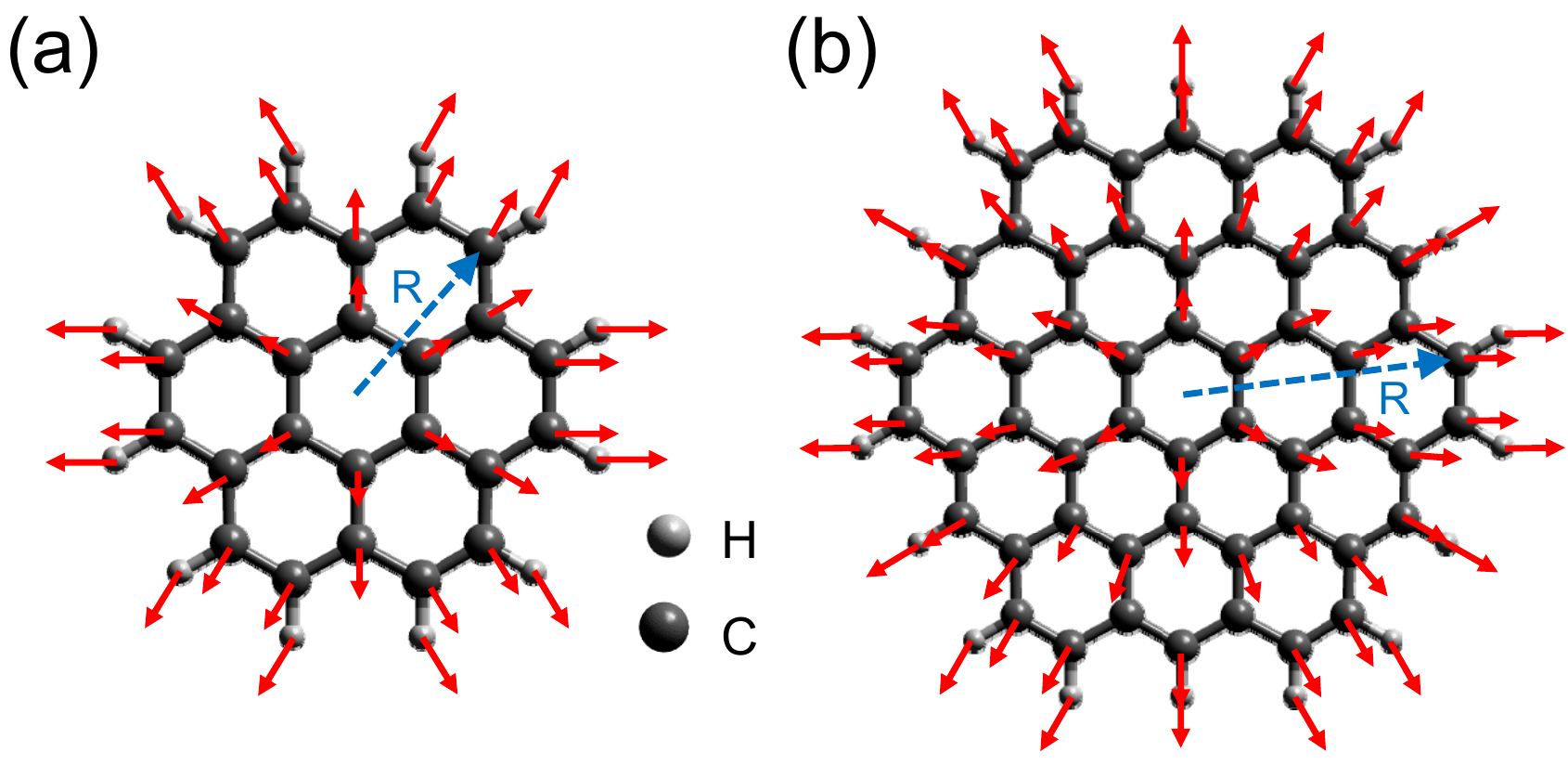}
\caption{Atomic structure of hydrogen-terminated circular graphene
discs with the radii %
(a) $R=3.75$~{\AA}, %
(b) $R=6.20$~{\AA}. %
Atomic motion during the radial breathing motion, based on DFT
calculations, is depicted by displacement vectors that are shown
by the red arrows.} %
\label{figS2}
\end{figure}

The structure of different hydrogen-terminated graphene discs is
represented by ball-and-stick models in Fig.~\ref{figS2}. The disc
radius $R$ is defined as the distance from the center to the
outermost carbon atom. We consider two discs, identical to those
described in Fig.~2(d) of the main manuscript, and display a snapshot
of the atomic motion during the radial breathing mode. We
found no notable difference in the character of the breathing modes
between results obtained by DFT and the REBOII force fields used
in the main manuscript.

%